# Control of particle trapping in a magnetoplasmonic nanopore

Nicolò Maccaferri,[1] Paolo Vavassori[2,3], and Denis Garoli[4,5]

[1]*Department of Physics and Materials Science, University of Luxembourg, 162a, avenue de la Faïencerie, L-1511 Luxembourg, Luxembourg*
[2]*CIC nanoGUNE BRTA, Tolosa Hiribidea, 76, E-20018 Donostia-San Sebastian, Spain*
[3]*IKERBASQUE, Basque Foundation for Science, Plaza Euskadi, 5, E-48009 Bilbao, Spain*
[4]*Faculty of Science and Technology Free University of Bozen, Piazza Università 5, 39100 Bolzano, Italy*
[5]*Istituto Italiano di Tecnologia, Via Morego 30, 16163 Genova, Italy*

**Plasmonic nanopores are extensively investigated as single molecules detectors. The main limitations in plasmonic nanopore technology are the too fast translocation velocity of the molecule through the pore and the consequent very short analysis times, as well as the possible instabilities due to local heating. The most interesting approach to control the translocation of molecules and enable longer acquisition times is represented by the ability to efficiently trap and tune the motion of nanoparticles that can be used to tag molecules. Here, we theoretically investigate the performance of a magneto-plasmonic nanopore prepared with a thin layer of cobalt sandwiched between two gold layers. A nanopore is then coupled with a translocating magnetic nanoparticle. By setting the magnetic configuration of the cobalt layer around the pore by an external magnetic field, it is possible to generate a nanoscale magnetic tweezer to trap the nanoparticle at a specific point. Considering a 10 nm magnetite nanoparticle we calculate a trapping force up to 28 pN, an order of magnitude above the force that can be obtained with standard optical or plasmonic trapping approaches. Moreover, the magnetic force pulls the nanoparticle in close contact with the plasmonic nanopore's wall, thus enabling the formation of a nanocavity enclosing a deeply sub-wavelength confined electromagnetic field with an average field intensity enhancement up to 230 at near-infrared wavelengths. The presented hybrid magneto-plasmonic system points towards a strategy to improve nanopore-based biosensors for single-molecule detection and potentially for analysis of various biomolecules.**

In the recent years, the engineering of electromagnetic (EM) fields at the nanoscale has become a subject of intense research. Plasmonics is an area of nanoscience dealing with electromagnetic coupled electronic oscillations (named Surface Plasmon Polaritons, SPPs) at the interface between metals and dielectric environments[1-5]. SPPs allow to confine EM field down to nanometer volumes, enabling several applications, from sensing and spectroscopy to energy harvesting and optoelectronics[6-12,14]. Among the others investigated phenomena, plasmonics has been proposed and demonstrated as a powerful tool to improve the performance of optical trapping technologies[15,16]. In fact, while standard methods use a focused laser beam to trap and manipulate objects with spatial resolution close to the wavelength of illumination, plasmonics enables to push this deeply subwavelength by at least one order of magnitude. Furthermore, plasmonic nanostructures have been used to manipulate dielectric and metallic nanoparticles down to few nm in diameter[17-23]. Recently it has been demonstrated that plasmonic architectures enable to trap single molecules with optical forces of few pN[24-27], thus opening up interesting opportunities in solid-state nanopore experiments. A nanopore is an aperture connecting two compartments and nowadays it is extensively investigated for sensing applications, with particular focus on single molecule detection[28,29]. The main limitations in solid-state nanopore technologies are the too-fast translocation time of the analyzed entities between the two compartments, as well as thermal instabilities due to local heating (enhanced Brownian motion, convective flows, turbulences), which prevent an efficient trapping of small objects for long times, thus hindering precise spectroscopy measurements. Typically driven by external electrical potential, a molecule or a nanoparticle interacts with the nanopore, which works as a sensor, in the time scale shorter than a microsecond. While this interaction time is enough for single entities detection, more advanced applications like single molecule sequencing are still a major challenge[28]. As recently demonstrated, a plasmonic nanopore can provide enhanced performances in optical spectroscopies[30,31], and enable a control in the velocity of translocation up to hundreds of milliseconds of both nanoparticles and single molecules such as DNA and proteins[32-34]. Although the highly focused light source used to excite the plasmonic nanopore can generate optical forces up to 3pN[35,36] and induce thermal effects that can facilitate the capture of the molecules, such thermal effects can be, at the same time, detrimental for the trapping[37-40], since localized optical heating and the consequent temperature gradient can increase the electrolyte conductivity and the thermal diffusion of the ions. The main consequence of these effects is the well-known thermophoretic effect[39]. In the continuous search of new approaches to improve the performances of solid-state nanopore devices, a new method to achieve strong trapping force by an external stimulus different from a laser excitation can be useful to overcome the practical limitations due to thermal instabilities during the trapping process.

In this Letter, we theoretically introduce and analyze a novel architecture where a plasmonic nanopore is modified by the presence of a thin layer of magnetic material to enable the trapping of magnetic nanoparticles by means of the generation and remote control of nanoscale magnetic tweezers by an external magnetic field. The concept is based on the recent demonstration of nanoscale trapping and manipulation of magnetic particles achieved through the magnetic field active control of the magnetic state of patterned magnetic thin films micro- and nanostructures[41]. In the present study, we were inspired by the robust trapping and programmable motion of magnetic nanoparticles achieved inside magnetic nano-rings by the application of time-variable external magnetic fields[42]. We consider a 200 nm large pore in a tri-layer Au[5nm]/Co[5nm]/Au[100nm] on top of a $Si_3N_4$ membrane 50 nm thick (Fig. 1a, left panel) and immersed in water (n=1.33), a standard device which is used in nanopore technology[18,19]. The diameter of the nanopore is chosen to make the system resonating at 780 nm (Fig. 1b), a typical wavelength used in fluorescence[29] and Raman[43-45] spectroscopies. A 10 nm diameter magnetic nanoparticle made of magnetite is allowed translocating through the nanopore. An in plane magnetic field of 1 kOe is applied temporarily to saturate Co layer along a desired direction and then reduced to 100 Oe to induce a magnetic configuration resulting in intense and localized magnetic traps (magnetic tweezers) at specific positions of the edge of the nanopore[46]. We analyze the generated force on the translocating nanoparticle and we show that the nanoparticle is pulled towards the nanopore's wall. In this configuration, a nanogap between the particle itself and the pore at specific positions can be generated (Figure 1a, right panel), thus creating a so called nanocavity between the pore's wall and the nanoparticle itself.

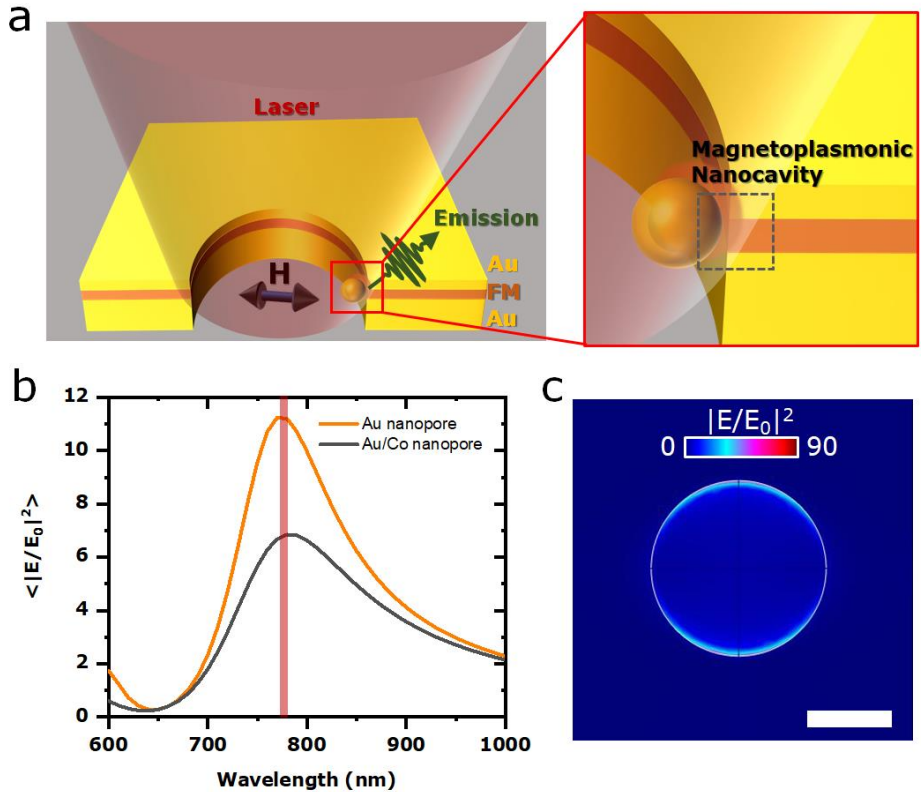

**Fig. 1.** (a) Left panel: geometry of the proposed experiment: a nanopore made of Au[5nm]/Co[5nm]/Au[100nm] immersed in water is illuminated with a laser at 780nm. A magnetic core-shell ($Fe_2O_3$-Au) nanoparticle is trapped due to the efficient force induced by the magnetic building block after the application of a static magnetic field **H**. Right panel: zoom on the nanoparticle-nanopore gap (magnetoplasmonic nanocavity). (b) Average field intensity enhancement in the nanopore in absence of the nanoparticle (the red line indicates the excitation wavelength). (c) Near-field intensity enhancement at 780 nm and at the upper Au/water interface. Scale bar: 100 nm.

We assume the incoming light to be tightly focused on the pore location (see Fig. 1a, left panel). This simple configuration is known to generate electric field confinement with low average intensity enhancement (i.e. $|E/E_0|^2 < 10$) in the nanopore volume (see Fig. 1c). Since our aim is to combine a magnetic field-induced control with plasmonic effects, a thin layer of cobalt is embedded in the Au film, 5 nm below the free surface of the film as depicted in Fig. 1a. As can be observed in Fig. 1b, compared with a nanopore made of pure gold, the presence of the thin Co layer is marginally perturbing the resonance position. On the other hand, the average field intensity enhancement $|E/E_0|^2$ in the nanopore is reduced by ~40% due to the higher losses related to the presence of cobalt in the system. Having briefly assessed the optical response of the nanopore, we now discuss the generation of controllable magnetic tweezers by controlling the magnetic configuration of the Co layer within the nanopore structure, which are utilized to stably trap a magnetic nanoparticle inside it and pull the particle close to the wall.

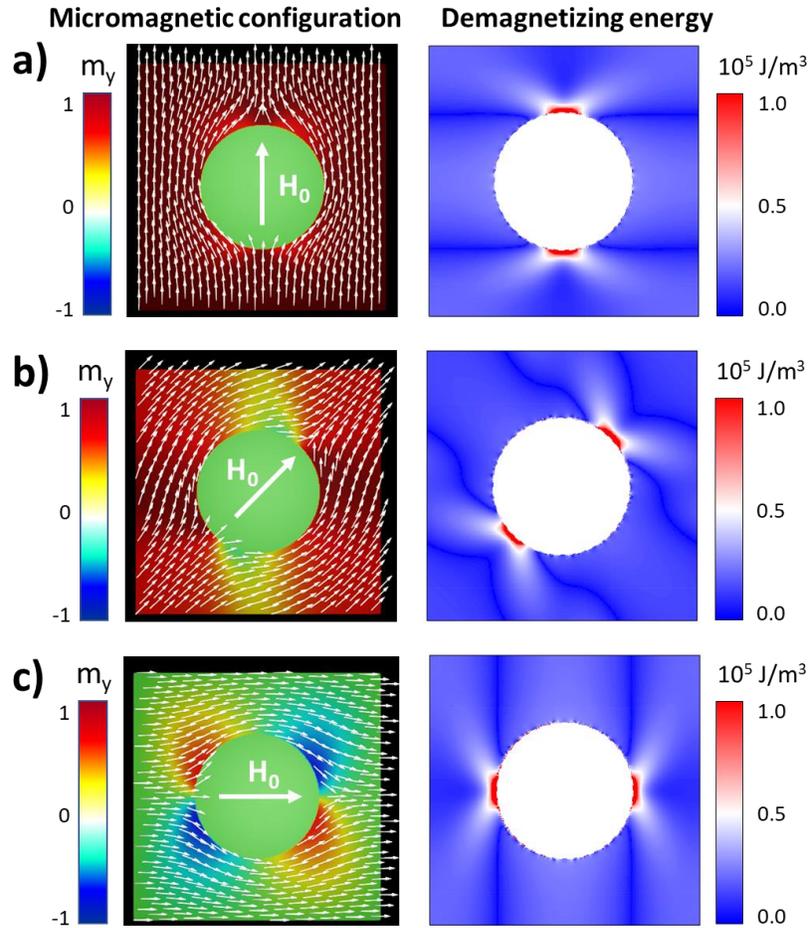

**Fig. 2.** Left-column panels show the relaxed micromagnetic configurations of the 5nm-thick Co layer upon the application of a remnant applied magnetic field $H_0$ of 100 Oe, which is rotated in plane by 90° as indicated in panels b) and c). Right-column panels show 2D maps of the demagnetizing energy density corresponding to the micromagnetic configurations in the left column.

The magnetic configuration in the 5-nm thick Co layer after applying a saturating magnetic field of 1 kOe along the y-axis and relaxing the magnetization in a remanent applied field $H_0$ of 100 Oe, still along the y-axis, is presented in Figure 2a (left panel). This was determined with micromagnetic simulations conducted by using mumax3[47] considering only the perforated Co layer. We considered a simulation unit of 400 nm x 400 nm with the 200 nm diameter pore in the center and applied periodic boundary conditions to account for an infinite film. The simulation unit was discretized in 256×256×2 finite difference cells of 1.56×1.56×2.5 nm$^3$, comparable to the exchange length of Co (2.6 nm).

The saturation magnetization and exchange stiffness used in the simulations are $M_s = 1400 \times 10^3$ A/m and $A = 30 \times 10^{-12}$ J/m respectively, corresponding to the tabulated values for the bulk material. The magnetocrystalline anisotropy is assumed to be vanishing small considering the typical polycrystalline structure of Co films deposited by thermal evaporation or sputtering on polycrystalline substrates (Au in this case). The right panel in Fig. 2a shows the 2D map of the demagnetizing energy density corresponding to the micromagnetic configuration in the left column. The small regions at the edge of the nanopore where the demagnetizing energy density is the highest, correspond to where large magnetic charges $\mathbf{M} \cdot \mathbf{n}$ and $-\nabla \cdot \mathbf{M}$ ($\mathbf{n}$ is the normal to the pore edge directed to the center of the pore), source of the intense and localized stray field (magnetic tweezers), are induced. The other panels b) and c) of Fig. 2 show the micromagnetic configurations upon rotation of the in-plane magnetic field $H_0$, resulting the controlled rotation of the tweezer along the pore circumference.

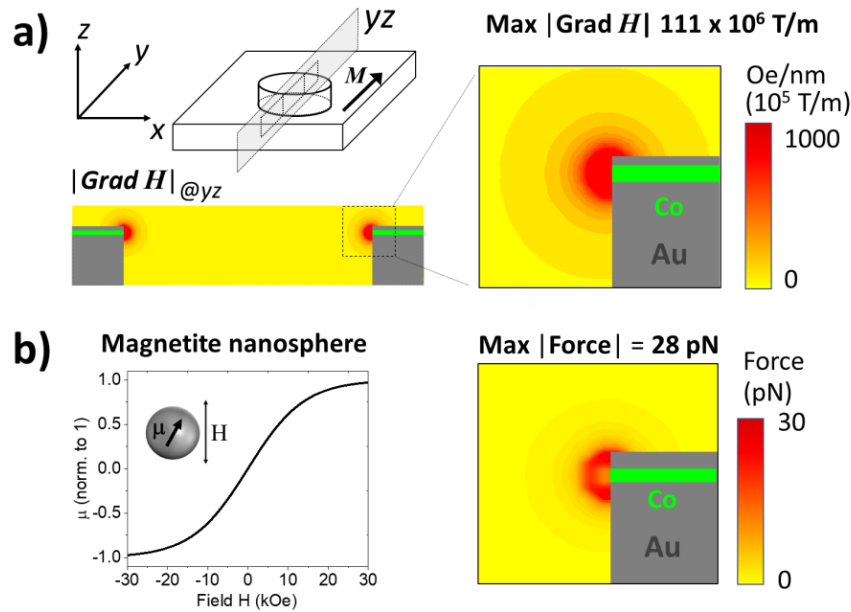

**Fig. 3.** Panel a) 2D map of the stray field gradient corresponding to the micromagnetic configuration in the Co layer shown in Fig. 2a. Panel b) calculated room temperature superparamagnetic response of a 10 nm diameter nanosphere of Magnetite (left) and 2D map of the corresponding force exerted on the nanosphere.

The simulations provide also the magnetic stray field **H** generated by the induced magnetic tweezers, from which the magnetic field gradient, **∇H**, is calculated. Figure 3a shows the 2D map of the calculated modulus of the magnetic field gradient in the plane-yz sketched in the figure. The 2D map displays an intense magnetic field gradient, reaching the considerable value of $10^8$ T/m, at the edge of the nanopore. The force acting on a magnetic nanoparticle depends on both the magnetic field gradient and the magnetic susceptibility of the nanoparticle[48]. The force expression is **F** = $\mu_0$(**μ·∇**)**H**; here **μ** = μ(H)**h**, where μ(H) is the magnetization curve of the nanoparticle and **h** is the unit vector parallel to the stray field **H**. The function μ(H) is shown in Figure 3b and is assumed to be a Langevin curve suitable for describing the room temperature superparamagnetic behavior of a 10 nm magnetite nanoparticle ($M_s = 450 \times 10^3$ A/m). The results of the calculations are summarized in the 2D map in Fig. 3b, where we plot a region in the xy-plane in proximity of one of the two magnetic tweezers and show that the magnetite nanoparticle is pulled against the upper edge of the pore with a maximum force of 28 pN (magnetic particle in contact with the pore's edge). The force reduces to 24 pN considering a gap of 2.5 nm between the magnetite nanoparticle and the pore edge to account for the thickness of an Au shell (2 nm) and the presence of a molecule (0.5 nm). It is worth mentioning here that, in the case of real magnetic nanoparticles, the susceptibility is even higher than the one we have estimated here[49]. Thus, the force can be in principle even bigger than the one we report here. As shown in Fig. 2, by rotating $H_0$ the position of the magnetic tweezers can be set along an arbitrary in-plane direction. Similarly to what discussed in Ref. [42], this interesting property can be used to generate a periodic movement of the nanoparticle inside the pore, e.g. a full rotation or an oscillation back and forth along the pore circumference. Considering the relative size of the nanoparticle with respect to the nanopore, this approach can be exploited to access different points around the pore's wall. For example, if the nanopore is functionalized with different molecules, multi-analytes spectroscopy can be performed just acting on the orientation of the applied magnetic field $H_0$, as depicted in Figure 2. Finally, it is also important to mention that a fast displacement of the magnetic tweezer can be utilized to release the nanoparticle[42].

After having demonstrated the possibility to generate trapping forces above 20 pN, we now focus on the estimation of the local electric field intensity enhancement for different distances between the magnetic nanoparticle and the nanopore internal side wall. We considered different geometries of the nanoparticle. In particular we focused on a core-shell nanoparticle with a core diameter of 10 nm as active magnetic medium (typically $Fe_2O_3$) and an Au shell of 2 nm, at three different distances from the nanopore's wall, thus for three different nanocavity dimensions. By changing the gap size from 2 nm to 0.5 nm, the local field intensity enhancement increases by a factor 6, from 20 to more than 130 (Figure 4a). Thus, an instability of 1 nm in controlling the particle-nanopore's wall distance by using an external magnetic field will not affect too much the local signal if we keep the particle nominally at 1 nm from the internal side wall of the nanopore. The possible drawback of this approach is that the particle dimensions seem to affect quite a lot the resonating wavelength of the cavity, which is, in this particular case, resonating around 860 nm, thus making very difficult to exploit the 780 nm wavelength, which can partially excite the nanocavity mode. A possible solution can be to design the nanopore diameter such that its resonance is blue-shifted towards lower wavelengths, so that when the nanopore is coupled with the nanoparticle the nanocavity mode resonates at 780 nm. Another solution can be maintaining the same particle diameter but changing the proportion between the materials composing the core and the shell building blocks. For a nanoparticle with a magnetic core of 8 nm and an Au shell of 4 nm, we can see that the advantage is two-fold. We have now a local field intensity enhancement of more than 230 at 780 nm and 0.5 nm particle-wall distance, thus providing enough signal coming from the nanocavity. In this latter case, namely, magnetic core 8 nm, Au shell 4 nm and a gap of 0.5 nm, the trapping force reduces to 15 pN so that the stability of the trapping mechanism is not dramatically affected.

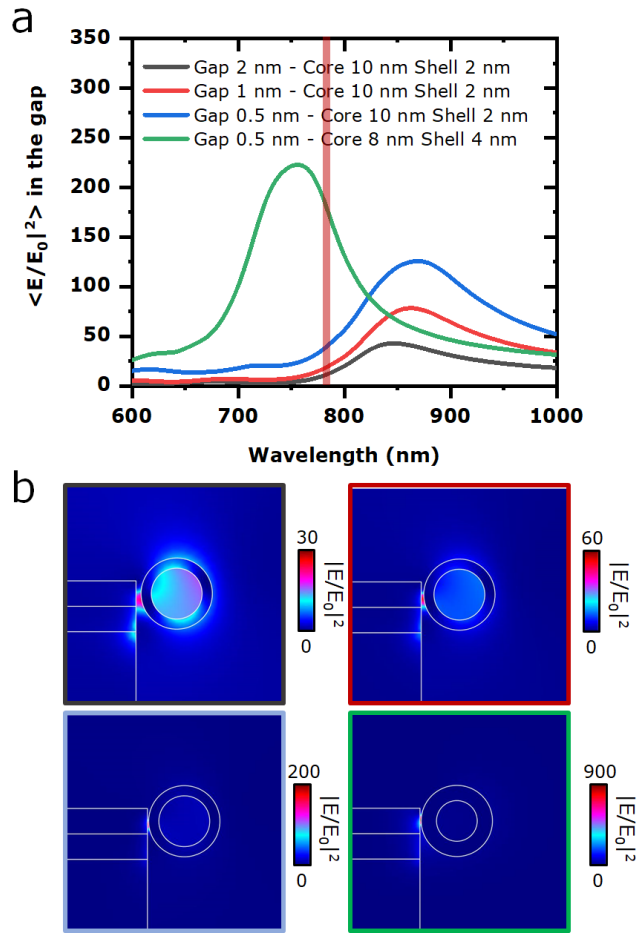

**Figure 4.** (a) Average field enhancement in the gap as function of the wavelength of the impinging light (red line indicates the excitation wavelength). (b) Near field intensity plots at 780 nm for various nanoparticle configurations/geometries.

In summary, we have studied a plasmonic nanopore capable of trapping a small magnetic nanoparticle as small as 10 nm thanks to intense nanometric magnetic tweezers generated, and remotely controlled, in a thin Co layer integrated in a Au nanopore membrane. The generated nanopore-nanoparticle coupled system enables the creation of a strong field confinement in a nanocavity between the particle and the pore's wall. The field intensity at the nanocavity can reach values up to more than 230 so potentially enabling single molecule spectroscopies. The dimensions used in the modeling and required here are feasible for nano-fabrication and integration in real world devices. It is worth noticing here that an important effect that can hinder the efficient trapping mechanism we have introduce in this work is represented by a temperature increasing upon light illumination. Nevertheless, typical fluences used in this type of experiments are of the order of 1-10 mJ/cm$^2$, and so we expect that the temperature might increase locally less than 10 °C, not enough to either demagnetize the Co layer or induce a thermal instability capable to displace the nanoparticle due to thermally-induced Brownian motion.

Interestingly, the higher losses related to the use of a hybrid magnetoplasmonic material instead of a pure noble metal material lead only to a slight modification of the local field intensity enhancement in the nanopore, and the significant trapping force (> 20pN) is an order of magnitude above the value that can be obtained in noble metal structures.

**Acknowledgements.** N.M. acknowledges support from the Luxembourg National Research Fund (Grant No. C19/MS/13624497 'ULTRON') and from the FEDER Program (Grant No. 2017-03-022-19 'Lux-Ultra-Fast'). P.V. acknowledges funding from the Spanish Ministry of Science and Innovation under the Maria de Maeztu Units of Excellence Programme (MDM-2016-0618) and the project RTI2018-094881-B-I00 (MICINN/FEDER).